\documentclass[12pt,thmsa]{article}
\begin{document}

\title{Collapsing shear-free perfect fluid spheres with heat flow}
\author{B.V.Ivanov \\
%EndAName
Institute for Nuclear Research and Nuclear Energy, \\
Bulgarian Academy of Science, \\
Tzarigradsko Shausse 72, Sofia 1784, Bulgaria}
\maketitle

\begin{abstract}
A global view is given upon the study of collapsing shear-free perfect fluid
spheres with heat flow. We apply a compact formalism, which simplifies the
isotropy condition and the condition for conformal flatness. The formulas
for the characteristics of the model are straight and tractable. This
formalism also presents the simplest possible version of the main junction
condition, demonstrated explicitly for conformally flat and geodesic
solutions. It gives the right functions to disentangle this condition into
well known differential equations like those of Abel, Riccati, Bernoulli and
the linear one. It yields an alternative derivation of the general solution
with functionally dependent metric components. We bring together the results
for static and time-dependent models to describe six generating functions of
the general solution to the isotropy equation. Their common features and
relations between them are elucidated$.$ A general formula for separable
solutions is given, incorporating collapse to a black hole or to a naked
singularity.
\end{abstract}

\section{Introduction}

Spherically symmetric radiating spacetimes are important in astrophysics for
modelling radiating stars and in cosmology. Gravitational collapse is a
highly dissipative process, required to account for the enormous binding
energy of the resulting object \cite{A27}. In the diffusion approximation
this is described by the heat flux. It allows to join the interior solution
to the Vaidya shining star exterior \cite{A11}. Radiating models are
necessary in cosmology to describe the formation of structure, evolution of
voids and the study of singularisties \cite{A30}.

Shear-free perfect fluids with heat flux are often studied in order to
simplify the calculations and allow realistic analytic solutions. Two of
their advantages are that there are just two metric components and their
space evolution is governed by the isotropy condition, which is an ordinary
second order linear differential equation in the radial variable \cite{A30},%
\cite{A16}. In the presence of heat flux the only non-trivial non-diagonal
component of the Einstein equations becomes an expression for it. The
vanishing of the heat flux implies a severe constraint, which transforms the
isotropy condition into a non-linear and highly complicated differential
equation with few explicit solutions \cite{A30},\cite{A1},\cite{A3},\cite
{book}. This situation persists even when anisotropy of the pressure is
allowed both for shear-free and geodesic fluids \cite{mine1},\cite{mine2}.

Many analytic solutions of the isotropy condition have been found. It has
been written in a compact form already in 1948 \cite{A1}. Published in an
obscure journal, this paper has been discussed nevertheless in some
monographs \cite{A30},\cite{book}. In spite of this, researchers prefer to
work directly with the metric coefficients, which complicates the
investigation. The isotropy condition is the same for static perfect fluid
models and for time-dependent ones. In the static case both the heat flux
and the off-diagonal component of the Einstein tensor vanish identically, so
there is no additional constraint on the isotropy condition. Time dependence
is obtained by promoting the integration constants into arbitrary functions
of time. As a result, there are two groups of authors - the static (S) and
the dynamical (D) group. Strangely enough, almost no interaction exists
between them. This problem becomes especially annoying when the general
solution of the isotropy equation is discussed. Recently, such a solution
was proposed by D authors \cite{A29}, using the Lie symmetries method. It
was treated as a class of solutions at first, but later this statement was
corrected \cite{A32}. However, generating functions have been found by the S
group as early as 1971 \cite{B5},\cite{B6}. Interestingly, the S authors
were not aware of previous research inside their group, so more such
functions appeared during the years \cite{B10},\cite{B12},\cite{B14},\cite
{B17},\cite{B18} (in the last reference a connection is made between it and
the previous one). Of course, only non-vanishing in the static limit
characteristics of the models were studied, like energy density, pressure
and the mass function.

While in the static case the main junction condition to the Schwarzschild
solution is the vanishing of the pressure on the surface, in the dynamical
case the correct joining was found in 1985 \cite{A11}. Like the isotropy
condition, this is an ordinary differential equation. The difference is that
it has only time derivatives and is non-linear. Once again, mainly its
formulation in terms of the metric coefficients was considered \cite{A25},%
\cite{A26},\cite{A28}. During the solving process it has been found that
some combinations simplify the computations.

One of the important questions of gravitational collapse is whether it ends
in a black hole or a naked singularity. A model with separable metric was
thoroughly studied \cite{A12},\cite{A13},\cite{A14} and it was found that a
black hole forms at the end. However, there exists a simple solution of the
same junction condition, when horizon never appears and the fate of the
collapsing matter is a naked singularity \cite{A24a}. Further, conformally
flat and geodesic models were studied extensively, but the fact that the
latter are a subclass of the first is not universally known.

In the present paper we address all these questions. In Sect.2 the field
equations are given as well as the definitions of important characteristics
of the fluid spheres like energy density, pressure, heat flux, expansion
scalar, mass function, the second Weyl invariant and the distribution of the
causal temperature. In Sect. 3 the junction conditions between the interior
and the exterior are presented and the definitions of quantities that lie or
depend on the surface are written. These are the surface luminosity,
redshift and temperature, the luminosity at infinity and the total energy
radiated during the collapse. It is stressed that the energy is stored in
the integration functions. The condition for the formation of a horizon and
consequently a black hole is also mentioned. Sect. 4 contains the compact
formulation of the isotropy equation (LG formalism) and its general
solution, obtained in six different ways. The relations between the
different generating functions are elucidated and the characteristics of the
model are expressed through some of them. In the next few sections several
classes of solutions are derived utilizing the LG formalism. These are
separable solutions in Sect. 5, conformally flat and geodesic solutions in
Sect. 6 and solutions with functional dependence between the metric
components in Sect. 7. For the sake of completeness we present
chronologically the remaining solutions known to us in Sect. 8. The last
section contains some conclusions.

\section{Field equations}

The collapse of a shear-free perfect fluid sphere is described by the
following metric in isotropic comoving coordinates 
\begin{equation}
ds^2=-A^2dt^2+B^2\left( dr^2+r^2d\Omega ^2\right) ,\quad d\Omega ^2=d\theta
^2+\sin ^2\theta d\varphi ^2,  \label{E1}
\end{equation}
where $A$ and $B$ depend on $r$ and $t$. The energy-momentum tensor for a
fluid undergoing dissipation in the form of heat flow reads \cite{A16},\cite
{A12} 
\begin{equation}
T_{ik}=\left( \mu +p\right) u_iu_k+pg_{ik}+q_iu_k+q_ku_i,  \label{E2}
\end{equation}
where $\mu $ is the energy density of the fluid, $p$ is the isotropic
pressure, $u_i$ is the four-velocity and $q_i$ is the radial heat flux
vector, which is orthogonal to $u_i$. In comoving coordinates 
\begin{equation}
u^i=A^{-1}\delta _0^i,\quad q^i=\left( 0,q,0,0\right) .  \label{E3}
\end{equation}
The non-trivial Einstein equations are 
\begin{equation}
8\pi \mu =\frac{3\dot B^2}{A^2B^2}-\frac 1{B^2}\left( \frac{2B^{\prime
\prime }}B-\frac{B^{\prime 2}}{B^2}+\frac{4B^{\prime }}{rB}\right) ,
\label{E4}
\end{equation}
\begin{equation}
8\pi p=\frac 1{A^2}\left( -\frac{2\ddot B}B-\frac{\dot B^2}{B^2}+\frac{2\dot
A\dot B}{AB}\right) +\frac 1{B^2}\left( \frac{B^{\prime 2}}{B^2}+\frac{%
2A^{\prime }B^{\prime }}{AB}+\frac{2A^{\prime }}{rA}+\frac{2B^{\prime }}{rB}%
\right) ,  \label{E5}
\end{equation}
\begin{equation}
8\pi p=\frac 1{A^2}\left( -\frac{2\ddot B}B-\frac{\dot B^2}{B^2}+\frac{2\dot
A\dot B}{AB}\right) +\frac 1{B^2}\left( -\frac{B^{\prime 2}}{B^2}+\frac{%
A^{\prime }}{rA}+\frac{B^{\prime }}{rB}+\frac{A^{\prime \prime }}A+\frac{%
B^{\prime \prime }}B\right) ,  \label{E6}
\end{equation}
\begin{equation}
8\pi q=\frac 2{B^2}\left( \frac{\dot B}{AB}\right) ^{\prime }.  \label{E7}
\end{equation}
Here the dot and the prime stand for time and radial derivatives
respectively. The rate of collapse $\Theta $ is given by \cite{A25} 
\begin{equation}
\Theta =\frac{3\dot B}{AB}  \label{E8}
\end{equation}
Let us use the variable $u=r^2$ and organize the time derivative terms in
terms of $\Theta $. The Einstein equations become 
\begin{equation}
8\pi \mu =\frac{\Theta ^2}3-\frac 4{B^2}\left( \frac{3B_u}B-\frac{uB_u^2}{B^2%
}+\frac{2uB_{uu}}B\right) ,  \label{E9}
\end{equation}
\begin{equation}
8\pi p=-\frac{\Theta ^2}3-\frac{2\dot \Theta }{3A}+\frac 4{B^2}\left[ \frac{%
uB_u}B\left( \frac{B_u}B+\frac{2A_u}A\right) +\frac{A_u}A+\frac{B_u}B\right]
,  \label{E10}
\end{equation}
\begin{equation}
8\pi q=\frac{4r\Theta _u}{3B^2},  \label{E11}
\end{equation}
\begin{equation}
\frac{2B_u^2}{B^2}+\frac{2A_uB_u}{AB}-\frac{A_{uu}}A-\frac{B_{uu}}B=0.
\label{E12}
\end{equation}
The last equation is the difference between Eq (5) and Eq (6) and represents
the isotropy of pressure. It contains no time derivatives and is an ordinary
second-order differential equation for $A,B$. Thus we have the freedom to
choose arbitrarily one of them and solve for the other. Time dependence
arises when the integration constants $C_i$ are promoted to integration
functions $C_i\left( t\right) $. Eqs (8-11) become expressions for $\Theta
,\mu ,p$ and $q$. The effective adiabatic index of the fluid $\Gamma =$d$\ln
p/$d$\ln \mu $ also can be found.

When the heat flow $q$ vanishes Eq (11) becomes another condition on $A,B$
and the combination with the isotropy condition leads to a non-linear
equation with few solutions \cite{A30},\cite{A1}, \cite{A3}, \cite{book}.
This situation remains also in the anisotropic case for shear-free or
geodesic collapsing spheres \cite{mine1}, \cite{mine2}.

The static case follows when the functions $C_i\left( t\right) $ become
constants again and $\Theta =0=$ $\dot A$. Then Eq (11) yields simply $q=0$
but the isotropy condition remains the same. Thus to every dynamical model
corresponds a static one.

An important characteristic in the general case is the mass function $m$ of
the fluid ball \cite{A25} 
\begin{equation}
\frac m{r^3}=\frac B2\left( \frac{\dot B^2}{A^2}-\frac{2B^{\prime }}{rB}-%
\frac{B^{\prime 2}}{B^2}\right) =\frac{B^3\Theta ^2}{18}-2B\left( \frac{B_u}%
B+\frac{uB_u^2}{B^2}\right) .  \label{E13}
\end{equation}

The conformal tensor has one essential component, which is given in an
invariant way by the second Weyl invariant $\Psi _2$. It can be determined
from the following formula, holding for anisotropic fluid spheres with heat
flow \cite{mine1} 
\begin{equation}
\frac m{R^3}=\frac{4\pi }3\left( \mu +p_t-p_r\right) -\Psi _2.  \label{E14}
\end{equation}
In the isotropic and shear-free case $R=rB$ and we have 
\begin{equation}
\Psi _2=\frac{8\pi \mu }6-\frac m{r^3B^3}.  \label{E15}
\end{equation}
Eqs (9,13) yield 
\begin{equation}
\Psi _2=\frac{4u}{3B^2}\left[ \left( \frac{B_u}B\right) ^2-\left( \frac{B_u}%
B\right) _u\right] .  \label{E16}
\end{equation}
The terms with time derivatives cancel and the equation for conformally flat
solutions $\Psi _2=0$ is an ordinary differential equation in $u$, like the
isotropy condition (12). Formula (14) has been derived originally in terms
of the electric part of the Weyl tensor $E$ \cite{Herrera}. Comparing both
formulas one comes to the conclusion that $E=3\Psi _2$.

Another characteristic of the collapsing sphere is the temperature
distribution $T$ among its volume. Unlike the previous quantities, it is
determined by a non-linear differential causal transport equation. For the
present metric it becomes \cite{A26} 
\begin{equation}
\tau \left( qB\right) ^{.}+qAB=-\frac{\kappa \left( AT\right) ^{\prime }}B,
\label{E17}
\end{equation}
where $\kappa $ is the thermal conductivity and $\tau $ is the relaxational
time-scale which gives rise to the causal behaviour of the theory. Both of
them depend on the temperature in general. A physically reasonable choice is
the transportation of energy by massless particles. Then Eq (17) becomes 
\begin{equation}
\frac \alpha 4X^{\prime }+qA^{4-\sigma }B^2X^{\sigma /4}+\beta \left(
qB\right) ^{.}A^3B=0,\quad X=\left( AT\right) ^4,  \label{E18}
\end{equation}
where $\alpha ,\beta ,\sigma $ are constants. In the non-causal case $\beta
=0$ it is easily solved. Integrable causal cases are $\sigma =0;2;4$ \cite
{A27},\cite{A26},\cite{A22},\cite{A23},\cite{A24}. Explicit solutions do not
alter the structure of this equation.

\section{Junction conditions}

The collapsing fluid lies within the sphere $\Sigma $ defined by $r=r_\Sigma 
$. The fluid is radiating, hence, the exterior is not vacuum, but the
outgoing Vaidya spacetime with the metric \cite{A16} 
\begin{equation}
ds^2=-\left( 1-\frac{2M\left( v\right) }\rho \right) dv^2-2dvd\rho +\rho
^2d\Omega ^2.  \label{E19}
\end{equation}
The junction conditions represent the continuity of the first and second
fundamental forms on $\Sigma $. This results in the relations \cite{A11} 
\begin{equation}
p=qB,\quad M\left( v\right) =m\left( r,t\right) ,\quad \rho \left( v\right)
=rB  \label{E20}
\end{equation}
which hold on the surface $\Sigma $. Here $M$ is the total mass of the
radiating sphere, $\rho _\Sigma $ is its radius as seen from outside and $v$
is the time of the distant observer. The first relation is a differential
equation containing derivatives with respect to $t$ only. Replacing Eqs
(10,11) and setting $r=r_\Sigma $ we get 
\begin{equation}
\Theta ^2+\frac{2\dot \Theta }A+\frac{4r\Theta _u}B=24\pi p_s,  \label{E21}
\end{equation}
where $p_s$ is the pressure of the corresponding static model.

Some of the characteristics of the model are defined on its surface. Such
are the surface luminosity $\Lambda _\Sigma $ and the redshift $z_\Sigma $ 
\cite{A16},\cite{Chan} 
\begin{equation}
\Lambda _\Sigma =\frac 23u_\Sigma ^{3/2}\left( B\Theta _u\right) _\Sigma
,\quad z_\Sigma =\left( 1+\frac{2uB_u}B+\frac{rB\Theta }3\right) _\Sigma
^{-1}-1.  \label{E22}
\end{equation}
The exterior time $v$ is related to the interior one $t$ as follows 
\begin{equation}
v=\int \left( 1+z_\Sigma \right) A_\Sigma dt.  \label{E23}
\end{equation}
The surface temperature of the star is 
\begin{equation}
T_\Sigma ^4=\frac{\Lambda _\Sigma }{4\pi \delta \rho _\Sigma ^2}=\left( 
\frac{r\Theta _u}{6\pi \delta B}\right) _\Sigma ,  \label{E24}
\end{equation}
where $\delta $ is some constant. The total luminosity for an observer at
rest at infinity reads 
\begin{equation}
\Lambda _\infty =-\frac{dM}{dv}=\frac{\Lambda _\Sigma }{\left( 1+z_\Sigma
\right) ^2}=\frac 23u_\Sigma ^{3/2}B\Theta _u\left( 1+\frac{2uB_u}B+\frac{%
rB\Theta }3\right) _\Sigma ^2,  \label{E25}
\end{equation}

The total energy radiated during the collapse of the fluid sphere follows
from the previous equation 
\begin{equation}
E_\infty =\int_{v_b}^{v_e}\Lambda _\infty dv=M\left( v_b\right) -M\left(
v_e\right) ,  \label{E26}
\end{equation}
where $v_b$ ($v_e$) is the exterior time of the collapse's start (end).
These correspond to $t_b$ ($t_e$) according to Eq (23). Eqs (13,20) show
that the radiated energy results from the change in the integration
functions $\bigtriangleup C_i=C_i\left( t_b\right) -C_i\left( t_e\right) $.
Thus the energy that a static model can give out is stored in its constants,
which are animated during the collapse and evolve under the single condition
given by Eq (21).

A separable solution was discussed in detail \cite{A12},\cite{A13},\cite{A14}%
. There $t_b=-\infty $ and the exterior solution at that moment is the
static exterior Schwarzschild solution in isotropic coordinates

\begin{equation}
ds^2=-\left( \frac{1-M_0/2\rho }{1+M_0/2\rho }\right) ^2dt^2+\left( 1+\frac{%
M_0}{2\rho }\right) ^4\left( d\rho ^2+\rho ^2d\Omega ^2\right) .  \label{E27}
\end{equation}
Here $M_0$ is the constant total mass. The sphere collapses until a black
hole is formed at $t_e=t_{BH}$. This happens when the coefficient $g_{00}$
in the exterior metric (19) vanishes at the surface $\Sigma $ and a horizon
appears. Eq (20) shows that the relation $2m=rB$ should be satisfied there.
It may be written with the help of Eq (13) as follows 
\begin{equation}
\left( 1+\frac{2uB_u}B+\frac{rB\Theta }3\right) _\Sigma \left( 1+\frac{2uB_u}%
B-\frac{rB\Theta }3\right) _\Sigma =0.  \label{E28}
\end{equation}
The first multiplier vanishes in order to satisfy this relation. Then Eqs
(22, 25) lead to the blowing up of the redshift and vanishing of the
luminosity at infinity.

Another separable solution with $t_b=-\infty $, however, never develops a
horizon \cite{A24a} and the collapse proceeds till all the mass is burnt
out, namely $M\left( t_e\right) =0$, which maximizes $E_\infty $. The final
state can also be any static model for anisotropic fluids with vanishing
radial pressure $p_r=0$ and no heat flow \cite{Joshi11}. In this case Eq
(21) is trivially satisfied.

\section{Six generating functions}

All of the previous formulas are based on the solution of Eq (12). The
search for its solutions spans an interval of 63 years, starting from 1948
when Kustaanheimo and Qvist wrote it in a very compact form \cite{A30},\cite
{A1},\cite{book}. Introducing instead of $A$ and $B$ the potentials $L=1/B$
and $G=A/B$ it becomes 
\begin{equation}
2GL_{uu}=LG_{uu},\quad B=1/L,\quad A=G/L.  \label{E29}
\end{equation}
This is a linear second-order differential equation. One can choose an
ansatz for $G$ and solve for $L$ or vice versa. However, a general solution
is hard to find. Choosing the function $K=L_u/L$ transforms Eq (29) into a
Riccati equation for $K$%
\begin{equation}
K_u+K^2-\frac{G_{uu}}{2G}=0,  \label{E30}
\end{equation}
which is first order, but still a general solution for any $G$ is not known.

This difficulty may be overcome if we choose one of the potentials as a
function of $AB$ or its $u$-derivative. For example let us take $A$ and $%
W=1/AB$. Then 
\begin{equation}
L=AW,\quad G=A^2W  \label{E31}
\end{equation}
and Eq (29) becomes 
\begin{equation}
\frac{A_u}A=\pm \sqrt{\frac{W_{uu}}{2W}},\quad B=1/AW.  \label{E32}
\end{equation}
It is readily integrable when an arbitrary $W$ is given \cite{A29}. 
\begin{equation}
A=A_0\left( t\right) \exp \pm \int \sqrt{\frac{W_{uu}}{2W}}du.  \label{E33}
\end{equation}
The function of integration $A_0\left( t\right) $ may be removed from $A$ by
a time change but it remains in the expression for $B$. In the above
reference the solution was presented as a result of Lie symmetries analyses,
together with other classes of solutions. Later it was emphasized that this
is a general solution \cite{A32} and in this way the potential of Msomi,
Govinder and Maharaj $W$ is a generating function for the metric $A,B$ and
all of the above characteristics of the model can be expressed through it.
In addition, $W$ together with Eqs (32-33) comprises the general solution of
the isotropy condition Eq (29). The potential $\hat W=W^{-1}=AB$ was also
discussed \cite{A29} with similar expressions for $A,B$.

The story does not begin here, however. As emphasized, the isotropy
condition holds also for the static case, which was studied extensively in
the past. Unfortunately, most of the authors worked in the so-called
curvature coordinates. Yet there is some amount of papers in isotropic
coordinates. Quite a few concrete solutions have been found and beside them
five other generating functions. As we pointed out in the introduction, the
work of this 'static' group of authors is completely unknown to the
'dynamical' group, which studied the time-dependent metric. The opposite is
also true. One of the purposes of the present paper is to present both the
static and dynamical results together, so that the future efforts may be
united and rediscoveries avoided. We shall derive the generating functions
from one another. Amazingly, their authors were unaware of the work of each
other with one small exception.

Thus, let us take instead of $W$ the potential 
\begin{equation}
U=-\frac{2W_u}W=2\left( \ln AB\right) _u,  \label{E34}
\end{equation}
\begin{equation}
U_u=\frac 12U^2-H^2,\quad H^2=\frac{4A_u^2}{A^2}.  \label{E35}
\end{equation}
In this form the isotropy condition was given by Kuchowicz \cite{B5},\cite
{B6},\cite{B7}. For $U$ this is a Riccati equation, but for $H$ is an
algebraic one and for $A$ is a simple linear equation. The metric is given
by 
\begin{equation}
A=\exp \frac 12\int Hdu,\quad B=B_0\left( t\right) \exp \frac 12\int \left(
U-H\right) du,  \label{E36}
\end{equation}
where $B_0$ is a function of integration. Thus $U$ is another generating
function. The change $H^2=UJ$ turns the Riccati equation for $U$ into a
Bernoulli equation, which is integrable, or into a simple algebraic equation
for $J$%
\begin{equation}
U_u=\frac 12U^2-JU.  \label{E37}
\end{equation}
Here one can choose either $U$ or $J$ as a generating function.

Let us replace next the potentials $U,H$ by $f,g$ according to the following
expressions 
\begin{equation}
U=\frac 1f,\quad H=\frac gf,  \label{E38}
\end{equation}
so that 
\begin{equation}
\frac 1f=2\left( \ln AB\right) _u,\quad g=\frac{\left( \ln A\right) _u}{%
\left( \ln AB\right) _u}.  \label{E39}
\end{equation}
Then the isotropy condition becomes simply 
\begin{equation}
f_u=g^2-\frac 12  \label{E40}
\end{equation}
and one can choose either $f$ or $g$ as a generating function. The other one
follows immediately, while the metric is given by 
\begin{equation}
A=\exp \frac 12\int \frac gfdu,\quad B=B_0\exp \frac 12\int \frac{1-g}fdu.
\label{E41}
\end{equation}
This third generating function was found by Goldman \cite{B10}, who also
gave some particular solutions. His work was corrected and further developed
by Knutsen \cite{B14}. The latter author expressed the characteristics of
the static model in term of the potentials, studied the energy conditions
and proposed another particular solution.

Next, let us take the potential 
\begin{equation}
\Phi =\ln AB=\int \frac{du}{2f}.  \label{E42}
\end{equation}
Then Eq (40) yields 
\begin{equation}
g=\varepsilon \frac{\sqrt{\Phi _u^2-\Phi _{uu}}}{\sqrt{2}\Phi _u},
\label{E43}
\end{equation}
where $\varepsilon =\pm 1$ and the metric is expressed through $\Phi $%
\begin{equation}
A=\exp \varepsilon \frac{\sqrt{2}}2\int \sqrt{\Phi _u^2-\Phi _{uu}}du,\
B=B_0\exp \left( \Phi -\varepsilon \frac{\sqrt{2}}2\int \sqrt{\Phi _u^2-\Phi
_{uu}}du\right) .  \label{E44}
\end{equation}
These are essentially the expressions of Ref.\cite{B18} and $\Phi $ is the
fourth generating potential, proposed by Lake.

The fifth one $Z$ was introduced by Rahman and Visser \cite{B17}. It is
obtained from the relation 
\begin{equation}
\Phi =2\int \frac Z{1-uZ}du  \label{E45}
\end{equation}
and is connected to the Goldman-Knutsen potentials by the equations

\begin{equation}
2f=\frac 1Z-u,\quad g^2=-\frac{Z_u}{2Z^2}.  \label{E46}
\end{equation}
Inserting them in Eq (41) we get for the metric 
\begin{equation}
A=\exp \pm \frac 1{\sqrt{2}}\int \frac{\sqrt{-Z_u}}{1-uZ}du,\quad
B=B_0A^{-1}\exp \int \frac Z{1-uZ}du.  \label{E47}
\end{equation}

Finally, we should mention the potentials introduced by Stewart \cite{B12} 
\begin{equation}
P=2r\left( \ln AB\right) _u,\quad S=2r\left( \ln \frac AB\right) _u,
\label{E48}
\end{equation}
which satisfy the equation 
\begin{equation}
2rP_u-\frac Pr-\frac 12P^2+SP+\frac 12S^2=0.  \label{E49}
\end{equation}
It can be shown that 
\begin{equation}
P=\frac rf,\quad S=\left( 2g-1\right) P  \label{E50}
\end{equation}
and then Eq (49) transforms into Eq (40). In this way $P$ is the sixth (and
the last known to us) generating function.

Eq (40) shows that to every function $f$ correspond two functions $\pm g$,
Suppose we have a solution $A_1,B_1$ with potentials $f_1,g_1$. Then there
is another solution with $f_2=f_1$ and $g_2=-g_1$. Eq (41) yields

\begin{equation}
A_2=A_1^{-1},\quad B_2=B_1A_1^2.  \label{E51}
\end{equation}
This is exactly the Buchdahl theorem \cite{B20} in the spherically symmetric
case.

Now let us give the characteristics of the fluid model in terms of $L$ and $%
G $: 
\begin{equation}
ds^2=L^{-2}\left( -G^2dt^2+dr^2+r^2d\Omega ^2\right) ,  \label{E52}
\end{equation}
\begin{equation}
\Theta =-\frac{3\dot L}G,\quad 8\pi q=\frac 43rL^2\Theta _u,  \label{E53}
\end{equation}
\begin{equation}
8\pi \mu =\frac{\Theta ^2}3+12L_u\left( L-uL_u\right) +8uLL_{uu},
\label{E54}
\end{equation}
\begin{equation}
8\pi p=-\frac{\Theta ^2}3-\frac{2\dot \Theta L}{3G}+\left( \frac{4G_uL}%
G-6L_u\right) \left( L-2uL_u\right) -2LL_u,  \label{E55}
\end{equation}
\begin{equation}
m=\frac{r^3}{L^3}\left[ \frac{\Theta ^2}{18}+2L_u\left( L-uL_u\right)
\right] ,\quad \Psi _2=\frac 43uLL_{uu},  \label{E56}
\end{equation}
\begin{equation}
\Lambda _\Sigma =\frac 23\left( \frac{r^3\Theta _u}L\right) _\Sigma ,\quad
T_\Sigma ^4=\frac 1{6\pi \delta }\left( r\Theta _uL\right) _\Sigma ,
\label{E57}
\end{equation}
\begin{equation}
z_\Sigma =\left( \frac{2uL_u+r\Theta /3}{L-2uL_u-r\Theta /3}\right) _\Sigma ,
\label{E58}
\end{equation}
\begin{equation}
\Lambda _\infty =\frac{\Lambda _\Sigma }{L_\Sigma ^2}\left( L-2uL_u-r\Theta
/3\right) _\Sigma ^2,  \label{E59}
\end{equation}
It is clear that $G$ appears only through $\Theta $, except in the pressure.
Note also the simple formula for the second Weyl invariant.

Finally, let us express the characteristics in terms of a generating
function. The most convenient are the Goldman-Knutsen potentials. $A$ and $B$
are found from Eq (41). The rest are 
\begin{equation}
\Theta =\frac 3{2A}\int \left( \frac{1-g}f\right) ^{.}du,  \label{E60}
\end{equation}
\begin{equation}
8\pi \mu =\frac{\Theta ^2}3-\frac 1{f^2B^2}\left[ \left( 1-g\right) \left(
6f+3u-ug-4ug^2\right) -4ufg_u\right] ,  \label{E61}
\end{equation}
\begin{equation}
8\pi p=-\frac{\Theta ^2}3-\frac{2\dot \Theta }{3A}+\frac 1{f^2B^2}\left[
2f+u\left( 1-g^2\right) \right] ,  \label{E62}
\end{equation}
\begin{equation}
\frac m{\left( rB\right) ^3}=\frac{\Theta ^2}{18}+\frac{g-1}{2f^2B^2}\left[
2f+u\left( 1-g\right) \right] ,  \label{E63}
\end{equation}
\begin{equation}
\Psi _2=\frac u{3f^2B^2}\left[ \left( 1-g\right) \left( 2+g-2g^2\right)
+2fg_u\right] ,  \label{E64}
\end{equation}
\begin{equation}
\Lambda _\infty =\frac{\Lambda _\Sigma }{\left( 1+z_\Sigma \right) ^2},\quad
z_\Sigma =\left( 1+\frac{1-g}fu+rB\Theta /3\right) _\Sigma ^{-1}-1,
\label{E65}
\end{equation}
while $q,\Lambda _\Sigma $ and $T_\Sigma $ are given by Eqs (11,22,24)
respectively. In the junction condition (21) one should put 
\begin{equation}
8\pi p_s=\frac 1{f^2B^2}\left[ 2f+u\left( 1-g^2\right) \right] ,  \label{E66}
\end{equation}
computed at the surface $\Sigma $. Here $g$ is given by Eq (41) and $f\left(
r,t\right) $ is an arbitrary function of $u$ and a number of functions $%
C_i\left( t\right) $. When the latter and $B_0\left( t\right) $ are
constant, $q,\Theta ,\Lambda _\Sigma ,\Lambda _\infty ,T_\Sigma ,T,p_\Sigma $
vanish, so that the model becomes static. Then the formulas for the energy
density, the pressure and the mass coincide with those of Knutsen \cite{B14}.

\section{Separable solutions}

Separable solutions have the following metric \cite{A6} 
\begin{equation}
A=j\left( t\right) \alpha \left( u\right) ,\quad B=h\left( t\right) \beta
\left( u\right) .  \label{E67}
\end{equation}
The function $j\left( t\right) $ may be set to $1$ by a time change. Then Eq
(12) shows that $\alpha $ and $\beta $ should satisfy the isotropy condition
as a static metric. Thus the generating functions are static, while $h\left(
t\right) =B_0\left( t\right) $. Replacing the above metric in the
expressions for the various characteristics we obtain 
\begin{equation}
\Theta =\frac{3\dot h}{\alpha h},\quad 8\pi q=-\frac{4r\alpha _u\dot h}{%
\alpha ^2\beta ^2h^3},  \label{E68}
\end{equation}
\begin{equation}
\Lambda _\Sigma =-2\dot h\left( \frac{r^3\alpha _u\beta }{\alpha ^2}\right)
_\Sigma ,\quad T_\Sigma ^4=-\frac{\dot h}{2\pi \delta h^2}\left( \frac{%
r\alpha _u}{\alpha ^2\beta }\right) _\Sigma ,  \label{E69}
\end{equation}
\begin{equation}
8\pi \mu =\frac{3\dot h^2}{\alpha ^2h^2}+\frac{8\pi \mu _s}{h^2},\quad 8\pi
p=-\frac{\dot h^2+2h\ddot h}{\alpha ^2h^2}+\frac{8\pi p_s}{h^2},  \label{E70}
\end{equation}
\begin{equation}
m=\frac{\left( r\beta \right) ^3}{2\alpha ^2}h\dot h^2+hm_s,\quad \Psi _2=%
\frac{\Psi _{2s}}{h^2},  \label{E71}
\end{equation}
\begin{equation}
\Lambda _\infty =\Lambda _\Sigma \left( 1+\frac{2u\beta _u}\beta -\frac{%
r\beta \dot h}\alpha \right) _\Sigma ^2,  \label{E72}
\end{equation}
\begin{equation}
z_\Sigma =\left( \frac{-2u\alpha \beta _u+r\beta ^2\dot h}{\alpha \beta
+2u\alpha \beta _u-r\beta ^2\dot h}\right) _\Sigma ,  \label{E73}
\end{equation}
where quantities with an index $s$ correspond to the static model with
metric $\alpha ,\beta $.

We suppose that the boundary of the static model is also at $r_\Sigma $, so
there $p_s=0$ and Eq (21) becomes 
\begin{equation}
2h\ddot h+\dot h^2-2a\dot h=0,\quad a=2\left( \frac{r\alpha _u}\beta \right)
_\Sigma .  \label{E74}
\end{equation}
This ordinary second-order differential equation governs the behaviour of $h$%
. Its first integral reads 
\begin{equation}
\dot h=\frac 2{\sqrt{h}}\left( a\sqrt{h}-b\right) ,  \label{E75}
\end{equation}
where $b$ is an integration constant. Integrating once more gives 
\begin{equation}
t-t_0=\frac h{2a}+\frac b{a^2}\sqrt{h}+\frac{b^2}{a^3}\ln |\sqrt{h}-\frac
ba|.  \label{E76}
\end{equation}
Here $t_0$ is a second integration constant. When $b\neq 0$ one can enforce
the equality $b=a$ by absorbing a constant in $\beta $ \cite{A12},\cite{A13},%
\cite{A14},\cite{A23},\cite{A24}. This solution was analysed in great detail
and leads to the formation of a black hole.

When $b=0$ the solution is $h=2a\left( t-t_0\right) $ \cite{A24a}. In this
reference this particular solution was given as a simple ansatz satisfying
Eq (74), the general solution being unavailable, according to the authors.
Here we have derived it from the general formalism. It is easy to see that $%
m $ and $rB$ are proportional to $t-t_0$, hence, their ratio is constant. At 
$\Sigma $%
\begin{equation}
\left( \frac{2m}{rB}\right) _\Sigma =2\left( \frac{8u^2\alpha _u^2}{\alpha ^2%
}+\frac{m_s}{r\beta }\right) _\Sigma .  \label{E77}
\end{equation}
This certainly can be made less than unity by choosing the constants of the
arbitrary static solution. Going back to the arguments that lead to Eq (28),
one comes to the conclusion that no horizon is formed during the process of
collapse. Its end is marked by $t_e=t_0$ when the mass burns out completely
and vanishes. The energy accumulated during the collapse is radiated at the
same rate. Both luminosities, the surface temperature and redshift are
constant, while $\mu ,p$ and $\Psi _2$ diverge as $\left( t-t_0\right)
^{-2}, $ $q\sim \left( t-t_0\right) ^{-3}$ and $\Theta \sim \left(
t-t_0\right) ^{-1}$. This may be an indication for the formation of a naked
singularity. In Ref. \cite{A24a} the special solution 
\begin{equation}
\alpha =1+cu,\quad \beta =1  \label{E78}
\end{equation}
was considered, which in addition is conformally flat. Solutions with no
horizon appear also in higher dimensional spacetimes \cite{A26a},\cite{A31}.

\section{Conformally flat solutions}

These solutions have $\Psi _2=0$ and a look at Eq (56) shows that the LG
formalism is the most appropriate for their study. Eqs (29,56) yield 
\begin{equation}
L_{uu}=0,\quad G_{uu}=0.  \label{E79}
\end{equation}
Integration produces four integration functions. In the general case they
are independent and Eq (52) indicates that in $G$ one of them may be set to
unity. Hence 
\begin{equation}
L=C_1\left( t\right) u+C_2\left( t\right) ,\quad G=u+C_3\left( t\right) .
\label{E80}
\end{equation}
The metric and the combination $AB$, which is the basis of the six
generating functions, become 
\begin{equation}
A=\frac{u+C_3}{C_1u+C_2},\quad B=\frac 1{C_1u+C_2},\quad AB=\frac{u+C_3}{%
\left( C_1u+C_2\right) ^2}.  \label{E81}
\end{equation}
The characteristics of the fluid model depend on the three functions $%
C_i\left( t\right) $. It is much simpler to use $C_1,L$ and $G$ instead. One
should note that 
\begin{equation}
L_u=C_1,\quad G_u=1,\quad \Theta _u=-\frac{3\dot C_1}G+\frac{3\dot L}{G^2}.
\label{E82}
\end{equation}
We can use Eqs (52-59) for the characteristics of the model, inserting in
them the above formulas. Eq (56) yields 
\begin{equation}
m=\frac{4\pi \mu ru}{3L^3},  \label{E83}
\end{equation}
which follows also from Eq (15) or the square of the conformal tensor \cite
{A25}. The time evolution of $C_1,L$ and $G$ is governed by the junction
condition Eq (21) on $\Sigma $, where $p_s$ is given by Eq (55) with $\Theta
=0$. Dropping for a while the index $\Sigma $, we obtain after a lengthy
calculation 
\[
2L\dot L\dot G+\left( 3\dot L^2-2L\ddot L\right) G+4rL\dot LG-4r\dot
C_1LG^2-4\left( L-2uC_1\right) LG^2+
\]
\begin{equation}
+\left( 8C_1L-12uC_1^2\right) G^3=0.  \label{E84}
\end{equation}

With respect to $G$ this is an Abel equation of the first kind 
\begin{equation}
A_1\dot G+A_2G+A_3G^2+A_4G^3=0.  \label{E85}
\end{equation}
It is not soluble analytically in general, but in some degenerate cases one
obtains \cite{A28} an algebraic equation (when $A_1=0$) or Bernoulli
equations (when $A_3$ or $A_4$ vanishes), which are integrable.

With respect to $C_1$ this is a Riccati equation 
\begin{equation}
R_1\dot C_1+R_2C_1+R_3C_1^2+R_4=0,  \label{E86}
\end{equation}
which may be transformed into a inhomogeneous second-order linear equation.
The general solution may be found if a particular one is known.

Finally, in order to elucidate the character of Eq (84) with respect to $L$
we make the replacement 
\begin{equation}
L=l^{-2},  \label{E87}
\end{equation}
which gives 
\[
G\ddot l-\left( \dot G+2rG\right) \dot l-G^2l-rG^2l^3\dot C_1+2G^2\left(
u+G\right) l^3C_1- 
\]
\begin{equation}
-3uG^3l^5C_1^2=0.  \label{E88}
\end{equation}
One can make this equation linear and homogenous in $l$ in two ways. First,
we simply put $C_1=0$ \cite{A25}, \cite{A26}. When $G$ is constant there are
three classes of solutions. In these references $G$ was taken in the form $%
G=C_3u+1$, which leads to more involved coefficients of the equations. They
hold for $u$ in general, but are needed for $u_\Sigma $ only. Another
possibility is to make $L\sim C_1$ This happens when $C_2=kC_1$ with $k$
some coefficient. The case of constant $G$ was studied \cite{A27}. In fact,
in these two cases $A$ and $B$ become separable and Eq (88) is the analogue
of the integrable Eq (74). Some other conformally flat and separable
solutions with a single, linear in time, integration function have been
discussed \cite{A24a},\cite{A18}.

A well known example of a static conformally flat solution is the interior
Schwarzschild metric. In isotropic coordinates it looks like \cite{B17}, 
\cite{B8} 
\begin{equation}
A=\frac{1+\frac{1+c_2}{c_2}\frac u{c_1}}{1+u/c_1},\quad B=\frac 1{1+u/c_1}
\label{E89}
\end{equation}
and has two constants. The limit $c_2\rightarrow \infty $ leads to the
Einstein universe, while $c_2=-1/2$ yields the De Sitter universe \cite{B17}%
. The $Z$ potentials of these solutions were given in the last reference. As
pointed out in the beginning, one can 'animate' these classical static
models by making the constants time-dependent. A model where they, in
addition, depend on each other was given by Kramer \cite{A20} and studied
later \cite{A21}. The junction condition becomes a complicated nonlinear
second-order differential equation, which surprisingly may be solved in
terms of a special function. Due to their simple structure, conformally flat
solutions were among the first to be discovered \cite{A30},\cite{A2a},\cite
{A8},\cite{A9},\cite{A10},\cite{A17},\cite{A19}. The subclass with $G=1$ was
studied too \cite{A7}.

An important class of solutions are the geodesic or non-accelerating ones,
which have $A=1$. This leads to $G=L$ and the metric becomes 
\begin{equation}
ds^2=-dt^2+L^{-2}\left( dr^2+r^2d\Omega ^2\right) .  \label{E90}
\end{equation}
Eq (29) gives $L_{uu}=0$ and Eq (56) shows that the geodesic solutions are a
subclass of the conformally flat solutions. We have for the expansion 
\begin{equation}
\Theta =-\frac{3\dot L}L,\quad \Theta _u=-3\left( \frac{C_1}L\right) ^{.}.
\label{E91}
\end{equation}
The characteristics of the model are obtained when these formulas are
inserted in Eqs (53-59).For example, the pressure reads 
\begin{equation}
8\pi p=-\frac{\Theta ^2}3-\frac{2\dot \Theta }3-4LC_1+4uC_1^2.  \label{E92}
\end{equation}
The condition $A=1$ simplifies the generating functions of geodesic models.
Thus $W=L$, $H=0$, $g=0$, $\Phi =-\ln L$, $Z=const$ and $P=-S$.

The junction condition Eq (21) becomes 
\begin{equation}
-4rL^2\dot C_1+4L\left( r\dot L+L^2\right) C_1-4uL^2C_1^2=2L\ddot L-5\dot
L^2,  \label{E93}
\end{equation}
taken on the surface $\Sigma $. It differs substantially from Eq (84),
although it still represents a Riccati equation for $C_1$ \cite{A28a}. When
some of the coefficients are made to vanish, a Bernoulli or a soluble
Riccati equation follows. The ansatz 
\begin{equation}
L=L_0\left( t+t_0\right) ^n,  \label{E94}
\end{equation}
where $L_0,t_0.n$ are constants, transforms Eq (93) into the linear
second-order equation of the confluent hypergeometric function. Elementary
solutions were found for $n=0;-2/3;-2$ \cite{A28a}. This method also
reproduces the solution with 
\begin{equation}
B=\frac{c_1}{2c_2}\left( \frac{1-c_2c_3\exp s}{1-uc_3\exp s}\right)
s^2,\quad s=\left( \frac{6t}{c_1}\right) ^{1/3},  \label{E95}
\end{equation}
where $c_i$ are constants. It has been discussed extensively in the past 
\cite{A22},\cite{A24}, \cite{A15},\cite{A15a},\cite{A20a}. Finally, it
should be mentioned that the study of collapsing shear-free perfect fluid
models with heat flow began with the well-known Robertson-Walker
cosmological model by promoting its constant $k$ to a function $k\left(
t\right) $ \cite{A30},\cite{A2a}, \cite{A17},\cite{A5},\cite{A19a}.

\section{Functional dependence between $A$ and $B$}

A class of solutions to the isotropy condition has the functional dependence 
$A\left( B\right) $. The latter is equivalent to $G\left( L\right) $. Then
Eq (29) becomes 
\begin{equation}
2GL_{uu}=G_{LL}LL_u^2+G_LLL_{uu},  \label{E96}
\end{equation}
which may be written as 
\begin{equation}
\frac{G_{LL}}{\frac{2G}L-G_L}L_u=\frac{L_{uu}}{L_u}.  \label{E97}
\end{equation}
Integrating once we obtain 
\begin{equation}
L_u=C_1\left( t\right) \exp \int \frac{G_{LL}}{\frac{2G}L-G_L}dL,
\label{E98}
\end{equation}
where $C_1\left( t\right) $ is an integration function. A second integration
gives 
\begin{equation}
\int \exp \left( -\int \frac{G_{LL}}{\frac{2G}L-G_L}dL\right) dL=C_1\left(
t\right) u+C_2\left( t\right) ,  \label{E99}
\end{equation}
$C_2\left( t\right) $ being a second integration function. Choosing an
explicit $G\left( L\right) $ one obtains after the integrations an explicit
or implicit expression for $L\left( t,u\right) $. The result for the
variables $A,L$ is similar 
\begin{equation}
\int \exp \left( -\int \frac{2A_L+LA_{LL}}{A-LA_L}dL\right) dL=C_1\left(
t\right) u+C_2\left( t\right) .  \label{E100}
\end{equation}
This formula was found long ago \cite{A9} and considered to be the general
solution of the isotropy equation. Formally this is true, because time plays
the role of a parameter and not a variable in it. Therefore, effectively, $%
A=A\left( u\right) $ and $B=B\left( u\right) $. Inverting the second
equality and replacing it in the first, one finds that for every solution
indirectly $A=A\left( L\right) $. Thus $L$ may be considered as another
generating function. However, the inversion and the double integration in
Eqs (99,100) make the procedure rather cumbersome and implicit. Formula
(100) was later rediscovered by the Lie symmetry method \cite{A29} and a few
examples were presented for illustration in both references. In the LG
formalism one of them is given by $G=L^3$. A brief calculation yields from
Eq (99) 
\begin{equation}
L=\left[ C_1\left( t\right) u+C_2\left( t\right) \right] ^{1/7}.
\label{E101}
\end{equation}
When $A$ is a function of $B$ (or $L$) such is the base for the generating
functions $AB$. Then Eq (39) shows that the Goldman-Knutsen potential $%
g=g\left( L\right) $. However, $f$ depends on $L$ and $L_u$ in the general
case.

A model of the same type was found in an attempt to generalize the exterior
Schwarzschild solution (27) \cite{A2},\cite{B3}. It has 
\begin{equation}
A=\frac{1-F}{1+F},\quad B=B_0\left( t\right) \left( 1+F\right) ^2,\quad F=%
\frac{c_1}{\left( 1+c_2u\right) ^{1/2}}.  \label{E102}
\end{equation}
Bayin \cite{B11} studied static solutions in isotropic coordinates of the
type 
\begin{equation}
A=A_0\phi ^{-c_1},\quad B=B_0\phi ^{c_2}  \label{E103}
\end{equation}
with 
\begin{equation}
\phi =c_3e^{c_4u},\quad \phi ^{1-c}=c_3u+c_4,  \label{E104}
\end{equation}
where $c=c\left( c_1,c_2\right) $ in a specified way. They also have $A$ and 
$B$ directly dependent on each other.

\section{Other solutions}

As stated before, static and time-dependent solutions are on equal footing
with respect to the isotropy condition. We present only one of the metric
functions in most cases, preserving the original notation. The constants
below are understood. Chronologically, the first solutions were given by
Narlikar et al in 1943 \cite{book},\cite{B6},\cite{B1}. They are static and
have 
\begin{equation}
L_I=C_1r^{1+n/2}+C_2r^{1-n/2},\quad L_{II}=C_3r^{-k/2},  \label{E105}
\end{equation}
where $n>0$, $0\geq k\geq -2$. There are three cases of $A$ for each $L$,
including the interior Schwarzschild metric (SIM). Nariai \cite{book},\cite
{B2} found five static solutions, one of them coinciding with $L_I$ from
above. The rest are 
\[
L_I^2=\left( a+bu\right) ^\alpha ,\quad L_{II}^2=u\left( a+b\ln r\right)
^2,\quad L_{III}^2=a\cos \left( b+cu\right) , 
\]
\begin{equation}
A_{IV}^2=\left( a+bu\right) ^\alpha .  \label{E106}
\end{equation}

In 1968 Strobel \cite{A30},\cite{A2a} gave a list of cases of $L_{uu}/L$ for
which a solution of Eq (29) can be found in the handbooks on differential
equations and two explicit solutions with $L_{uu}=0$. This reference
remained unnoticed. Kuchowicz \cite{book},\cite{B5},\cite{B6},\cite{B7},\cite
{B4} gave a host of new static solutions, using his generating function. In
view of their great number we direct the reader to the original references.
Goldman \cite{B10} studied three explicit static examples of his potential $%
g $%
\begin{equation}
g_I=\frac a{1-bu},\quad g_{II}=\frac 1{\sqrt{2}}\coth \left( a-bu\right)
,\quad g_{III}=\cosh \left( a+bu\right) .  \label{E107}
\end{equation}
Stewart \cite{book},\cite{B12} applied the Buchdahl theorem to SIM to obtain
a new static solution with 
\begin{equation}
A^2=c\left( 1+au\right) ^2\left( 1-bu\right) ^{-2}.  \label{E108}
\end{equation}
Sanyal and Ray \cite{A9} gave their Case 1 dynamical solution 
\begin{equation}
A=C\left( t\right) u+D\left( t\right)  \label{E109}
\end{equation}
as a complementary one to their general solution (100). Modak \cite{A10}
proposed a time-dependent metric, which coincides with the animated fourth
Nariai solution. Pant and Sah \cite{B13} studied in detail the static model
with 
\begin{equation}
A=A_0\frac{1-k\delta }{1+k\delta },\quad B=\frac{\left( 1+k\delta \right) ^2%
}{1+u/a^2},\quad \delta \left( u\right) =\frac{\left( 1+u/a^2\right) ^{1/2}}{%
\left( 1+bu/a^2\right) ^{1/2}},  \label{E110}
\end{equation}
which is a generalisation of the Buchdahl solution (102) for $b\neq 0$.

Deng in 1989 invented a powerful method (Deng's ladder) \cite{A30},\cite{A17}
for generating an infinite chain of more and more complicated solutions,
varying with time, from a simple seed $A_1$. One finds next the general form
of $L_1$, takes it as a seed, finds the general $A_2$ and so on. He
delivered the most general conformally flat solution and some others.
Banerjee et al \cite{A18a} gave two dynamical solutions 
\begin{equation}
A_I=\frac{T\left( t\right) z^{1/2}-\alpha }{T\left( t\right) z^{1/2}+\alpha }%
,\quad A_{II}=\frac z{T\left( t\right) \left( z+a/T\left( t\right) \right)
},\quad z=1+\eta \left( t\right) u.  \label{E111}
\end{equation}

In 1990 Knutsen \cite{B14} corrected and extended the Goldman potential
method by studying the characteristics, the physical plausibility and the
dynamical stability of the static models in terms of the Goldman-Knutsen
generating function. He discussed the third Goldman solution and proposed a
simpler one with $g=au+b$.

Another static solution was given by Burlankov \cite{B15} 
\begin{equation}
L^2=a\left( \frac{u+b-\sqrt{3}/2}{u+b+\sqrt{3}/2}\right) ^{\sqrt{3}}.
\label{112}
\end{equation}
Recently, Pant et al \cite{B19} studied in detail the static metric 
\begin{equation}
A=\frac{\cos \sqrt{2}b\left( d-u\right) }{\sin \left( a+bu\right) },\quad
B=\cos ec^2\left( a+bu\right) .  \label{E113}
\end{equation}
Finally, Msomi, Govinder and Maharaj \cite{A29}, with the help of Lie
symmetry analysis, found five transformations, leading to new solutions from
old ones and introduced their generating function $W$.

\section{Conclusions}

We have given a global view upon the study of collapsing shear-free perfect
fluid spheres with heat flow. The application of the LG formalism has been
advocated throughout the present paper. It provides a very compact
formulation of the isotropy condition (12), namely Eq (29), and a very
simple expression for $\Psi _2$ - Eq (56). The formulas for the other
characteristics, Eqs (52-59), are also straight and tractable. Eq (56)
clearly shows why the condition for conformal flatness is so similar to the
isotropy condition.

The LG formalism also presents the simplest possible version of the junction
condition. This has been demonstrated explicitly for conformally flat and
for geodesic solutions. It gives the right functions to disentangle this
condition into well known differential equations like the Abel equation, the
Riccati equation, the Bernoulli equation or the linear one. This formalism
yields an alternative derivation of the general solution when the metric
components are functionally dependent.

We have also discussed an unified study of separable solutions by
incorporating the simple linear in time ansatz into the general formula for
the solution of the junction condition (76).

One of the main objectives of the paper is to bring together the results of
the static and dynamical group of authors, not only in the chronology of
particular solutions, but in the discovery of generating functions. The
recent proposition of the generating potential $W$ \cite{A29} has prompted
the search for similar functions, mainly in the work of the static group. A
bunch of five generating potentials has been found, any of which provides
the complete solution of Eq (29). Their common feature is the presence of
the basic form $AB.$ Its use reduces the LG equation to a first order or an
algebraic one, depending on which potential of the pair is chosen as a seed.
It seems that the Goldman-Knutsen generating function satisfies the simplest
equation and a future task may be to continue the studies of Knutsen upon
characteristics, pertinent to the dynamical models. Obviously, new
generating potentials may be proposed by taking other functions of $AB$ or
its $u$-derivative. In view of this we hope that the enumeration of the
existing generating functions, undertaken here, will prevent wasting of time
in rediscoveries in the future.

Finally, putting in order the four-dimensional case will help to lessen the
efforts in investigating shear-free radiating collapse in higher dimensions 
\cite{A32},\cite{A31}.


\begin{thebibliography}{99}
\bibitem{A27}  Herrera, L., Di Prisco A., Ospino, J.: Phys. Rev. D \textbf{74%
}, 044001 (2006)

\bibitem{A11}  Santos, N.O.: Mon. Not. R. Astron. Soc. \textbf{216}, 403
(1985)

\bibitem{A30}  Krasinski, A.: Inhomogeneous Cosmological Models. Cambridge
University Press, Cambridge (1997)

\bibitem{A16}  Bonnor, W.B., De Oliveira, A.K.G., Santos, N.O.: Phys. Rep. 
\textbf{5}, 269 (1989)

\bibitem{A1}  Kustanheimo, P., Qvist, B.: Comm. Phys. Math. Helsingf. 
\textbf{13}, 16 (1948)

\bibitem{A3}  Glass, E.N.: J. Math. Phys. \textbf{20}, 1508 (1979)

\bibitem{book}  Stephani, H., Kramer, D., Maccalum, M., Hoenselaers, C.,
Herlt, E.: Exact Solutions to Einstein's Field Equations. Cambridge
University Press, Cambridge (2003)

\bibitem{mine1}  Ivanov, B.V.: Int. J. Mod. Phys. A \textbf{25}, 3975 (2010)

\bibitem{mine2}  Ivanov, B.V.: Int. J. Mod. Phys. D \textbf{20}, 319 (2011)

\bibitem{A29}  Msomi, A.M., Govinder, K.S., Maharaj, S.D.: Gen. Relativ.
Gravit. \textbf{43}, 1685 (2011)

\bibitem{A32}  Msomi, A.M., Govinder, K.S., Maharaj, S.D.: Int. J. Theor.
Phys. \textbf{51}, 1290 (2012)

\bibitem{B5}  Kuchowicz, B.: Phys. Lett. A \textbf{35}, 223 (1971)

\bibitem{B6}  Kuchowicz, B.: Acta Phys. Pol. B \textbf{3}, 209 (1972)

\bibitem{B10}  Goldman, S.P.: Astrophys. J. \textbf{226}, 1079 (1978)

\bibitem{B12}  Stewart, B.V.: J. Phys. A \textbf{15}, 1799 (1982)

\bibitem{B14}  Knutsen, H.: Gen. Relativ. Gravit. \textbf{23}, 843 (1991)

\bibitem{B17}  Rahman, S., Visser, M.: Class. Quantum Grav. \textbf{19}, 935
(2002)

\bibitem{B18}  Lake, K.: Phys. Rev. D \textbf{67}, 104015 (2003)

\bibitem{A25}  Herrera, L., Le Denmat, G., Santos, N.O., Wang, A.: Int. J.
Mod. Phys. D \textbf{13}, 583 (2004)

\bibitem{A26}  Maharaj, S.D., Govender, M.: Int. J. Mod. Phys. D \textbf{14}%
, 667 (2005)

\bibitem{A28}  Mishtry, S.S., Maharaj, S.D., Leach, P.G.L.: Math. Meth.
Appl. Sci. \textbf{31}, 363 (2008)

\bibitem{A12}  De Oliveira, A.K.G., Santos, N.O., Kolassis, C.A.: Mon. Not.
R. Astron. Soc. \textbf{216}, 1001 (1985)

\bibitem{A13}  De Oliveira, A.K.G., De F. Pacheco, J.A., Santos, N.O.: Mon.
Not. R. Astron. Soc. \textbf{220}, 405 (1986)

\bibitem{A14}  De Oliveira, A.K.G., Kolassis, C.A., Santos, N.O.: Mon. Not.
R. Astron. Soc. \textbf{231}, 1011 (1988)

\bibitem{A24a}  Banerjee, A., Chatterjee, S., Dadhich, N.: Mod. Phys. Lett.
A \textbf{17}, 2335 (2002)

\bibitem{Herrera}  Herrera, L., Ospino, J., Di Prisco, A., Fuenmayor, E.,
Troconis, O.: Phys. Rev. D \textbf{79}, 064025 (2009)

\bibitem{A22}  Govender, M., Maharaj, S.D., Maartens, R.: Class. Quantum
Grav. \textbf{15}, 323 (1998)

\bibitem{A23}  Govender, M., Maartens, R., Maharaj, S.D.: Mon. Not. R.
Astron. Soc. \textbf{310}, 557 (1999)

\bibitem{A24}  Govinder, K.S., Govender, M.: Phys. Lett. A \textbf{283}, 71
(2001)

\bibitem{Chan}  Pinheiro, G., Chan, R.: Gen. Relativ. Gravit. \textbf{40},
2149 (2008)

\bibitem{Joshi11}  Joshi, P.S., Malafarina, D., Narayan, R.: Class. Quantum
Grav. \textbf{28}, 235018 (2011)

\bibitem{B7}  Kuchowicz, B.: Acta Phys. Pol. B \textbf{4}, 415 (1973)

\bibitem{B20}  Buchdahl, H.A.: Aust. J. Phys. \textbf{9}, 13 (1956)

\bibitem{A6}  Glass, E.N.: Phys. Lett. A \textbf{86}, 351 (1981)

\bibitem{A26a}  Banerjee, A., Chatterjee, S.: Astrophys. Space Sci. \textbf{%
299}, 219 (2005)

\bibitem{A31}  Govinder, K.S., Govender, M.: Gen. Relativ. Gravit. \textbf{44%
}, 147 (2012)

\bibitem{A18}  Banerjee, A., Dutta Choudhury, S.B., Bhui, B.K.: Phys. Rev. D 
\textbf{40}, 670 (1989)

\bibitem{B8}  G\"urses, M., G\"ursey, Y.: Nuovo Cim. B \textbf{25}, 786
(1975)

\bibitem{A20}  Kramer, D.: J. Math. Phys. \textbf{33}, 1458 (1992)

\bibitem{A21}  Maharaj, S.D., Govender, M.: Aust. J. Phys. \textbf{50}, 959
(1997)

\bibitem{A2a}  Strobel, H.: Wiss. Z. Friedrich Schiller Univ. Jena. Math.
Naturw. Reihe \textbf{17}, 195 (1968)

\bibitem{A8}  Maiti, S.R.: Phys. Rev. D \textbf{25}, 2518 (1982)

\bibitem{A9}  Sanyal, A.K., Ray, D.: J. Math. Phys. \textbf{25}, 1975 (1984)

\bibitem{A10}  Modak, B.: J. Astrophys. Astr. \textbf{5}, 317 (1984)

\bibitem{A17}  Deng, Y.: Gen. Relativ. Gravit. \textbf{21}, 503 (1989)

\bibitem{A19}  Deng, Y., Mannheim, P.D.: Phys. Rev. D \textbf{42}, 371 (1990)

\bibitem{A7}  Som, M.M., Santos, N.O.: Phys. Lett. A \textbf{87}, 89 (1981)

\bibitem{A28a}  Thirukkanesh, S., Maharaj, S.D.: J. Math. Phys. \textbf{50},
022502 (2009)

\bibitem{A15}  Kolassis, C.A., Santos, N.O., Tsoubelis, D.: Astrophys. J. 
\textbf{327}, 755 (1988)

\bibitem{A15a}  Chan, R., Lemos, J., Santos, N.O., De F. Pacheco, J.A.:
Astrophys. J. \textbf{342}, 976 (1989)

\bibitem{A20a}  Grammenos, T.: Astrophys. Space Sci. \textbf{211}, 31 (1994)

\bibitem{A5}  Bergmann, O.: Phys. Lett. A \textbf{82}, 383 (1981)

\bibitem{A19a}  Jiang, S.: J. Math. Phys. \textbf{33}, 3503 (1992)

\bibitem{A2}  Nariai, H.: Prog. Theor. Phys. \textbf{38}, 92 (1967)

\bibitem{B3}  Buchdahl, H.A.: Astrophys. J. \textbf{140}, 1512 (1964)

\bibitem{B11}  Bayin, S.: Phys. Rev. D \textbf{18}, 2745 (1978)

\bibitem{B1}  Narlikar, V.V., Patwardhan, G.K., Vaidya, P.C.: Proc. Nat.
Inst. Sci. India \textbf{9}, 229 (1943)

\bibitem{B2}  Nariai, H.: Sci. Rep. Tohoku Univ. \textbf{34}, 160 (1950)

\bibitem{B4}  Kuchowicz, B.: Ind. J. Pure Appl. Math. \textbf{2}, 297 (1971)

\bibitem{B13}  Pant, D.N., Sah, A.: Phys. Rev. D \textbf{32}, 1358 (1985)

\bibitem{A18a}  Banerjee, A., Dutta Choudhury, S.B., Bhui, B.K.: Pramana 
\textbf{34}, 397 (1990)

\bibitem{B15}  Burlankov, D.E.: Theor. Math. Phys. \textbf{95}, 455 (1993)

\bibitem{B19}  Pant, N., Mehta, R.N., Pant, M.J.: Astrophys. Space Sci. 
\textbf{330}, 353 (2010)
\end{thebibliography}
\end{document}